\documentclass{article}
\usepackage{ijcai22}

\usepackage{times}
\usepackage{soul}
\PassOptionsToPackage{hyphens}{url}
\usepackage[hidelinks]{hyperref}
\usepackage[hyphenbreaks]{breakurl}
\usepackage[utf8]{inputenc}
\usepackage[small]{caption}
\usepackage{graphicx}
\usepackage{amsmath}
\usepackage{amsthm}
\usepackage{booktabs}
\usepackage{algorithm}
\usepackage{algpseudocode}
\usepackage{soul}
\usepackage{amsfonts,color}
\newcommand{\mc}{\mathcal}

\usepackage{caption}
\usepackage{subcaption}
\usepackage{comment}

\usepackage[english]{babel}
\usepackage{blindtext}

\usepackage{newfloat}
\usepackage{listings}
\usepackage{amsmath,amsthm,amsfonts,amssymb}

\DeclareMathOperator{\ud}{UD}
\DeclareMathOperator{\udh}{\widehat{UD}}
\DeclareMathOperator{\ug}{U\mathcal{G}}
\DeclareMathOperator{\tlg}{\tilde{g}}

\DeclareMathOperator{\gd}{GD}
\DeclareMathOperator{\dad}{\mathcal{G}D}
\DeclareMathOperator{\dt}{DT}

\usepackage{booktabs}

\floatstyle{ruled}
\newfloat{listing}{tb}{lst}{}
\floatname{listing}{Listing}

\title{Using Constraint Programming and Graph Representation Learning for Generating Interpretable Cloud Security Policies}

\author{
Mikhail Kazdagli$^{1,2}$\and
Mohit Tiwari$^{1,2}$\and
Akshat Kumar$^{1,3}$\\
\affiliations
$^1$Symmetry Systems\\
$^2$The University of Texas at Austin\\
$^3$Singapore Management University\\
\emails
\{mikhail, mohit, akshat.kumar\}@symmetry-systems.com
}

\usepackage{bibentry}

\newcommand{\sysname}{IAMAX}

\newcommand{\arxiv}[1]{}

\begin{document}

\maketitle

\begin{abstract}
Modern software systems rely on mining insights from business sensitive data stored in public clouds.
A data breach usually incurs significant (monetary) loss for a commercial organization.
Conceptually, cloud security heavily relies on Identity Access Management (IAM) policies that IT admins need to properly configure and periodically update.
Security negligence and human errors often lead to misconfiguring IAM policies which may open a backdoor for attackers. To address these challenges, \textit{first}, we develop a novel framework that encodes generating \textit{optimal} IAM policies using constraint programming (CP). We identify reducing \textit{dormant permissions} of cloud users as an optimality criterion, which intuitively implies minimizing unnecessary datastore access permissions. \textit{Second}, to make IAM policies interpretable, we use graph representation learning applied to historical access patterns of users to augment our CP model with \textit{similarity} constraints: similar users should be grouped together and share common IAM policies.
\textit{Third}, we describe multiple attack models and show that our optimized IAM policies significantly reduce the impact of security attacks using real data from 8 commercial organizations, and synthetic instances.
\end{abstract}

\section{Introduction}
\label{sec:intro}

Cloud computing has recently become the dominant computing paradigm which provides the flexibility of on-demand compute, and reduced cost due to economies of scale. However, such benefits come at a cost---private and business sensitive data is stored on public clouds. Managing identity access policies (IAM), which intuitively means deciding which user should have access to which datastore, for public clouds such as Amazon AWS~\cite{aws_iam}, is complex even for small and medium-sized companies with few hundred users and datastores~\cite{aws_iam_hard}. IAM policies are configured by IT admins of the organization who often do not have access to automated decision support tools for IAM policy optimization. As a result, due to the inherent complexity of IAM policy optimization, security negligence, and human errors may often lead to misconfigured IAM policies and result in data breaches~\cite{aqua_cloud_risks,breach1,linkedin_breach,fb_breach}.


\paragraph{Related work.} There has been prior work in AI and ML on securing the sensitive data in a cloud. Traditionally security researchers focus on developing dynamic behavioral detectors that analyze system behavior using (un)supervised ML methods at different levels - system and API calls~\cite{Canali2012,droid_mat}, hardware signals~\cite{Demme2013,sherlock}, network traffic~\cite{kitsune_nids,Sommer2010,network_intrusion_detection}, and domain reputations~\cite{gossip_kruegel,made_malicious_domains}. 
Such methods often suffer from raising unacceptably high volume of false positives~\cite{nodoze} when analyzing large amount of system events even if the false positive rate is very low~\cite{fp_security}. Moreover, ML-based detectors are susceptible to adversarial attacks~\cite{nlp_adversarial_attacks,neural_cleanse,adversarial_examples}. Our proposed method, \sysname, avoids such shortcomings by securing an organization's cloud infrastructure by hardening IAM policies, which are the first line of defense, rather than detecting attackers inside the cloud. Even after a user's account is compromised, our optimized IAM policies minimize the leak of sensitive data. 



There has been a history of applying tools from AI planning for security in the context of penetration testing (or pentesting)~\cite{SarrauteBH12,Hoffmann15,ShmaryahuS0S18}. In pentesting, automated tools based on planning are used to identify vulnerabilities in the network by launching controlled attacks under a variety of settings, such as fully or partially observable network settings. Once vulnerabilities are identified, they can be patched by network administrators. Our proposed work is different from such pentesting methods as our target is to \textit{design} the IAM policies (analogous to network design) from grounds up so that opportunities for catastrophic attacks (where compromising very few users gives hackers access to majority of datastores) are minimized. Furthermore, our approach is customized for public clouds and their security configurations, which is a different setting than the standard pentesting~\cite{ShmaryahuS0S18}.

\paragraph{Contributions.} Our main contributions are as follows. \textit{First}, we formalize the problem of IAM policy optimization for public clouds using the constraint programming (CP) framework~\cite{cp}, identfying core objectives and constraints. We highlight the key role that \textit{dormant permissions} play in defining secure IAM policies. \textit{Second}, given that organizations that we have worked with do not provide the identities and job roles of its cloud users due to privacy concerns, we use graph neural network~\cite{gcn,graph_sage} to learn embeddings for users and datastores based on the information flow between them. These embeddings are used for defining constraints that make IAM policies interpretable. \textit{Third}, we describe multiple attack models, based on randomly  compromising $k$ users, and adversially compromising those users that lead to worst case outcome. We test our optimized IAM policies on real cloud infrastructures of 8 medium size commercial companies and several realistic synthetic instances generated using the properties of real world data sets, and show the significant potential of our approach to secure cloud infrastructures by reducing dormant permissions significantly. 
We released the code and the data sets used in our experiments\footnote{\url{https://github.com/mikhail247/IAMAX}}.

\arxiv{\vspace{-0.1in}}
\section{IAM Policies and Cloud Computing}
\label{sec:iam}


An IAM policy is the fundamental security primitive in any cloud. Though our real world data sets were collected across multiple cloud platforms, we use Amazon AWS IAM policies as an example. 
An IAM policy is expressed using a declarative policy language that defines what operations (\textbf{``Action"}) are (dis)allowed (controlled by the \textbf{``Effect"} field) on a resource (\textbf{``Resource"} field). Figure~\ref{fig:iam_policy} shows an oversimplified IAM policy to be attached to an AWS identity (e.g. a user, a role) or a \textit{group} of identities. It stipulates that the identity that the policy is attached to is allowed to perform the operation ``s3:ListObjects" (read the bucket content) on the S3 bucket ``arn:aws:s3:::bucket-name".


\arxiv{\vskip{-2pt}}
\paragraph{Overpriviliged security policies.} To make the number of policies manageable, IT admins group identities together into permission groups and/or roles (on AWS) and attach IAM policies to them. We call such user aggregations \textbf{\textit{data access groups}}. Typically, each data access group carries a certain semantic meaning (e.g. web developers).
A policy assigned to a data access group grants access to all cloud resources that individual identities in the group access to.
Such an approach creates \textit{over-privileged} identities by granting each identity access to cloud resources that are used by some identities in a group~\cite{aqua_cloud_risks,verizon_dbir}. We call permissions that identities are granted, but never use, \textit{\textbf{dormant permissions}}. Dormant permissions violate the fundamental security principle - the \textit{principle of least privilege}~\cite{least_privilege} that stipulates that every user and an automated service should operate using the least set of privileges necessary to complete the job. 
Dormant permissions allow attackers after compromising an identity's credentials to not only get access to cloud resources that identity actively uses, but also get access to additional resources that the compromised identity has access to because of dormant permissions.

IAM policies are not necessarily poorly designed from the beginning - they are likely to deteriorate over time as organization evolves (e.g. new automated services get developed, employees move between teams, etc). Policies rarely get updated to reduce the number of dormant permissions, thus causing the set of dormant permissions to expand over time which gives an advantage to attackers.

\begin{figure}[tbp]
 {  
    \textbf{``Statement"}: [\{ \\
    \textbf{``Effect"}: ``Allow", \\
    \textbf{``Action"}: ``s3:ListObjects", \\
    \textbf{``Resource"}: ``arn:aws:s3:::bucket-name" \}] \\
}
\arxiv{\vspace{-0.2in}}
\caption{An example of an AWS IAM policy}
\label{fig:iam_policy}
\end{figure}

\begin{figure}[tbp]
\centering
\includegraphics[trim=0cm 3.35cm 0cm 4.9cm, clip=True, width=0.48\textwidth]{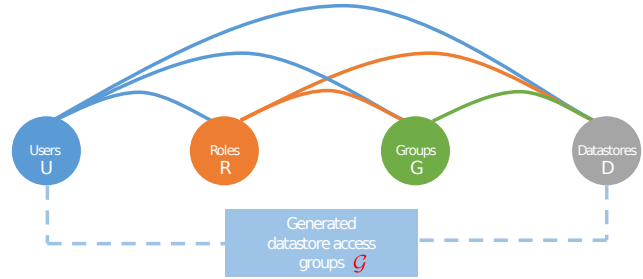}\par
\arxiv{\vspace{-0.1in}}
\caption{Schematic representation of the way users access datastores in AWS cloud. For simplicity we refer to both Roles and Groups as existing data access groups.}
\label{fig:aws_user_datastore}
\end{figure}

\arxiv{\vskip{-0.1in}}
\paragraph{Our approach.} We minimize the amount of dormant permissions to reduce damage caused by potential attacks.
We intentionally refrain from considering a trivial solution - block a user from accessing unused datastores.
Though such an approach may work, but in practice it is not being used because it would be hard for IT admins to maintain up-to-date per-user policies (specific accesses that should be allowed/blocked per user) especially when users transition between teams within an organization.

Figure~\ref{fig:aws_user_datastore} shows the schema denoting how users are partitioned into \textit{roles} and \textit{groups} and how access to different datastores is determined based on permissions granted to roles, groups and users themselves. Our approach to minimizing dormant permissions is to create a set of additional \textit{datastore access groups} $\mathcal{G}$ (shown in figure~\ref{fig:aws_user_datastore}). 
Solid edges represent existing IAM policies, while the dashed edges represent generated IAM policies. Conceptually, dashed edges are supposed to replace solid edges and thus remove dormant permissions. Both types of edges depict mapping of users to datastore access groups, which are named as roles and groups by cloud platforms, and mapping of those groups to datastores.
Using CP, we optimize: (1) partitioning of users into different generated groups $g\in \mc{G}$, and (2) what datastores are accessible by each group $g\in \mc{G}$. 
From an implementation perspective, the policy generated by the CP solver is concatenated with original over-permissive policies via \textit{logical AND} operation using cloud IAM policy management tools, thereby reducing dormant permissions. We did not edit existing IAM policies to avoid introducing unintentional errors due to high complexity of the policy language.

\section{Optimizing IAM Policies}
\label{sec:model}


We start by introducing different objects and relationship among them (as depicted in Figure~\ref{fig:aws_user_datastore}). These objects and relationships will define the variables, constraints and objective of the IAM policy optimization problem. Set $U$ denotes the set of users; $D$ denotes the set of datastores; $G$ denotes the set of existing data access groups; set $\mc{G}$ denotes the generated access groups as noted in Section~\ref{sec:iam}. Set ${T}$ denotes data types in a datastore according to the organization's data classification scheme (e.g, social security number, email, credit card number among others). Real world data sets used in our experiments contain at most 10 data types.
Such objects within an organization are extracted from mining its cloud infrastructure. The number of datastore access groups ($|\mc{G}|$) is a configurable parameter that intuitively defines the flexibility we have in reconfiguring the IAM policies (will be discussed later).
We deliberately avoid making the number of datastore access groups ($|\mc{G}|$)  to be a variable to reduce model complexity and thus increase scalability. 

\paragraph{Users and existing groups.} As noted in Section~\ref{sec:iam}, users are end users (either humans or services) that need to access different datastores.
Users may directly access a datastore, however, usually they either assume a role or inherit permissions from a permission group. 
Original IAM policies include 13--150 users, 90--1792 roles, and 5--833 permission groups.
Without loss of generality, we denote both roles and permission groups as `existing data access groups' $G$. Thus, users are mapped to different groups $G$, which are mapped to datastores.
We mine all such links, and create following relationships.

\paragraph{Users and datastores.} Based on existing permissions (users-roles-groups-datastores), we can also compute \textit{all} the datastores $d\in D$ that a user $u\in U$ can potentially access. This is stored in variable $\udh(u, d)\in \{0, 1\}$. These variables are constants as they are based on pre-defined permissions in an organization's cloud.

\paragraph{Users and accessed datastores.} Using an organization's historical data, we can determine which datastores have been accessed by different users. Let $\ud(u, d) \in \{0, 1\}$ denote if user $u\in U$ has accessed datastore $d\in D$ in the past. We shall require any refactored IAM policy to preserve access of different users to the datastores they have accessed in the past.

\paragraph{Groups and datastores.} Let $\gd(g, d)\in \{0, 1\}$ denote if group $g\in G$ provides access to the datastore $d\in D$. This is a constant mapping.

\paragraph{Users and datastore access groups.} We create $|\mc{G}|$ datastore access profiles that allow us to further control user access to datastores. 
Let variable $\ug(u, \tlg)\in \{0, 1\}$ denote whether user $u\in U$ belongs to the access group $\tlg\in \mc{G}$.

\paragraph{Datastore access groups and datastores.} Let variable $\dad(\tlg, d)\in \{0, 1\}$ denote whether datastore access group $\tlg\in \mc{G}$ provides access to the datastore $d\in D$.

\paragraph{Datastores and data types.} Let $\dt(d, t)\in \{0, 1\}$ denote whether datastore $d$ contains data of type $t\in T$. This is a constant mapping.

\paragraph{Constraints.} We now describe the relationship between different variables that define the IAM policy optimization problem. We next show for a given relationship among users, groups and datastore access groups, how to compute how many total datastores a user $u$ can access, and based on that compute \textit{redundant} datastores a user has access to. These redundant permissions constitute the so-called \textit{dormant permissions} that we want to minimize.

We create a variable $\widetilde{\ud}(u, d) \in \{0, 1\}$ to denote if user $u$ can access datastore $d$ \textit{after} we incorporate additional permissions and relations from the datastore access groups $\mc{G}$. Constraints on it are the following:
\paragraph{Frequently accessed datastores must still be accessible.} {$$\widetilde{\ud}(u, d) \geq \ud(u, d) \; \forall u\in U, \; d\in D$$}

\paragraph{Computing $\widetilde{\ud}(u, d)$.} A user $u$ can only access datastore $d$ iff the following conditions hold:
\begin{itemize}
    \item User $u$ is part of a datastore access group $\tlg$ and $\tlg$ has permission to access the datastore $d$ \textit{AND}
    \item The current permissions based on user-role-groups-datastore also allow user $u$ to access $d$
\end{itemize}
{\small 
\begin{align}
    \widetilde{\ud}(u, d)= \text{OR}\Big( \ug(u, \tlg)\bigwedge \dad(\tlg, d) \forall \tlg\in\mc{G}\Big) \bigwedge \udh(u, d) \nonumber 
\end{align}}
where OR stands for the logical \textit{or} operator.

\paragraph{Data type constraints.} A data type constraint stipulates that a user should not get an access to a completely new  data type  that they never worked before with. We make an assumption that a user accesses all data types that are in the datastore $d$ (if the user is allowed to access $d$).
More fine-grained formulation of this constraint would require operating at an object level (e.g. directory/table level in a datastore), but 
this would significantly increase the number of variables in the problem and make it computationally intractable.

The data type constraint can be formulated as:
\arxiv{\vskip{-2pt}}
{
\begin{align}
\text{OR}\Big( \ud(u, d)\bigwedge \dt(d, t)\; \forall d\in {D}\Big) \geq \nonumber \\ 
\text{OR}\Big( \widetilde{\ud}(u, d)\bigwedge \dt(d, t)\; \forall d\in {D}\Big) \forall u\in U, t\in T
\end{align}}

\subsection{Constraint Program for Reducing Dormant Permissions}
Based on the variables and relationships among them, we now describe the constraint program (CP) that optimizes IAM policies by minimizing the dormant permissions. Formally, the reduction in the number of user-datastore permissions in the refactored IAM policy is:
\arxiv{\vskip{-8pt}}
{
\arxiv{\small}
\begin{align}
    \sum_{u, d} \udh(u, d) - \sum_{u, d} \widetilde{\ud}(u, d)) \label{eq:dark}
\end{align}}
Maximizing the above objective would minimize the second term in the objective (first term is constant based on the existing IAM policy). Therefore, users in the refactored policy would have access to as few datastores as necessary given the problem constraints, and number of access groups $|\mc{G}|$. Thus, it will also reduce dormant permissions, and result in a \textit{least-privilege} IAM policy.

\paragraph{Penalty on dormant permissions per user.} The objective~\eqref{eq:dark} minimizes the total number of dormant permissions in the system. However, we also want to limit the maximum number of dormant permissions per user for increased robustness to attacks.
We formulate such constrains as a soft constraints, violation of which incurs a penalty of ${\psi_i}(v)$ computed as below:
{\arxiv{\small}
\begin{align}
    & v_u = \sum_{d} \widetilde{\ud}(u, d) - (1.0 + \epsilon) \ud(u, d) \forall u\in U \\
    &{\psi_{u} = \max (v_u, \gamma v_u) \forall u\in U}
\end{align}}
\arxiv{\vspace{-0.2in}}

where $\epsilon$ is a hyper-parameter in our model, which should be tuned by a domain expert according to the amount of risk they are willing to take.
If the parameter is set to 0, then our CP formulation will softly penalize any user that has non-zero amount of dormant permissions. 
Hence, if $\epsilon>0$, then the model does not penalize users with less than $\epsilon$ fraction of dormant permissions. 
Based on the empirical analysis of IAM policies, we believe that allowing a small fraction of dormant permissions (e.g. 15\%) increases policy interpretability 
and is unlikely to significantly affect security aspect of the problem.
The parameter $\gamma$ penalizes such soft constraint violations more harshly if it is more than 1.
We can also put per user dormant permissions as constraints, however in practice it made the program infeasible.
The overall CP formulation is given in Table~\ref{tab:cp}.

	\begin{table}[t!]
	\resizebox{3.35in}{!}{
	\begin{tabular}{|c|}
	\hline
	\begin{minipage}{0.49\textwidth}
	{\arxiv{\small}
	 \begingroup
         \addtolength{\jot}{0pt}
\begin{align}
    \max \Big(&\sum_{u, d} \udh(u, d) - \sum_{u, d} \widetilde{\ud}(u, d)\Big)  - \sum_{u} \psi_{u}   \\
    & \widetilde{\ud}(u, d) \geq \ud(u, d) \; \forall u\in U, \; d\in D \\
    & \widetilde{\ud}(u, d)= \text{OR} \Big( \ug(u, \tlg)\bigwedge \dad(\tlg, d)\; \forall \tlg\in\mc{G}\Big) \nonumber \\
    &\hspace{70pt}\bigwedge \udh(u, d) \; \forall u\in U, \; d\in D \label{eq:perm} \\
    &\text{OR}\Big( \ud(u, d)\bigwedge \dt(d, t)\; \forall d\in {D}\Big) \geq \nonumber \\ 
&\text{OR}\Big( \widetilde{\ud}(u, d)\bigwedge \dt(d, t)\; \forall d\in {D}\Big) \forall u\in U, t\in T\\
  & v_u = \sum_{d} \widetilde{\ud}(u, d) - (1.0 + \epsilon) \ud(u, d) \forall u\in U \\
    &{\psi_{u} = \max (v_u, \gamma v_u) \forall u\in U}\\
    &  \dad(\tlg, d), \ug(u, \tlg) \in \{0, 1\}\; \forall u, \tlg, d\\
    \nonumber 
\end{align}
\endgroup }
	\end{minipage} \\
	\hline
	\end{tabular}}
	\arxiv{\vspace{-6pt}}
	\caption{\small Constraint program for minimizing dormant permissions}
 	\arxiv{\vspace{-5pt}}
	\label{tab:cp}
	\end{table}


\subsection{Group Homogeneity Constraints}
\label{sec:gh}

We further enhance the basic CP formulation (Table~\ref{tab:cp}) by including group homogeneity constraints. Intuitively, all users in a group $\tlg \in \mc{G}$ should be \textit{similar} w.r.t.$\!$ a job role in an organization to make the solution more explainable and interpretable to IT admins. Moreover, heterogeneous groups may increase an impact of a potential attack: similar users are usually susceptible to the same attack vector (e.g. phishing, waterhole, etc). If they are spread across multiple groups, then attackers may get additional advantage if the number of residual dormant permissions across all those groups is larger than in the case when similar users are grouped together.

We assume a precomputed pairwise function ${\alpha_{user}}$ defined on $U$x$U$ denoting dissimilarity between users. Section~\ref{sec:gnn} shows how this function can be computed using graph neural networks (GNNs) and the data that we mine from an organization's cloud infrastructure. 
We add the homogeneity constraints as below. If a user $u$ is mapped to datastore access group $\tlg$ (i.e., $\ug(u, \tlg)=1$), then we use the shorthand $u\in \tlg$ for exposition clarity.
\begin{align}
\max \big\{ {\alpha_{user}}(u_1, u_2) \forall u_1, u_2\in \tlg \big\} \leq \alpha , \forall \tlg \in \mc{G} \label{eq:diversity}
\end{align}
where $\alpha$ is a diversity threshold that we show in Section~\ref{sec:gnn} how to compute from the data. 

\paragraph{Constraint generation.} Constraints~\eqref{eq:diversity} create memory issues when incorporated in a single program as quadratic number of terms are there in the number of users in a group $\tlg$. Also, they often make the CP problem infeasible. Therefore, we follow a constraint generation approach~\cite{Ben-Ameur2006} where we first solve the program in Table~\ref{tab:cp} with no homogeneity constraints. We then follow an iterative constraint generation procedure to add back most violated homogeneity constraints and solve the program again. 

\paragraph{Separation oracle.} At every iteration for each datastore access group $\tlg \in \mc{G}$, we identify user pairs $(u_1, u_2) \text{ s.t. } \ug(u_1, \tlg)=\ug(u_2, \tlg)=1$, and $\alpha_{user}(u_1, u_2) > \alpha$. 
We then add the constraint $\small \ug(u_1, \tlg) + \ug(u_2, \tlg) \leq 1 \;\; \forall \tlg \in \mathcal{G}$ to the CP and solve again.
This process continues until all the homogeneity constraints are satisfied or we get an infeasible program, in which case we output the solution generated in the last iteration. It may happen that no solution may exist that satisfies all the homogeneity constraints (based on the threshold $\alpha$). In such a case, the constraint generation approach provides the solution with most violated constraints being satisfied. 



\arxiv{\vskip{-2pt}}
\section{Graph Representation Learning}
\label{sec:gnn}



\paragraph{Incorporating GNN into CP.} Interpretable policies require group homogeneity with respect to users' job roles. However, such information is not shared with us due to high business sensitivity. Group homogeneity constraints are defined in terms of the user dissimilarity function ${\alpha_{user}}$ and the diversity threshold $\alpha$ (Section~\ref{sec:gh}). We use a graph neural network (GNN) to approximate both parameters and share them with the CP solver.

\paragraph{User behavioral graph (UBG).} Intuitively, job roles can be inferred from users' interaction with a cloud environment (e.g. executed operations, accessed cloud resources). We represent records of all executed cloud operations over the 6-12 months period as a weighted heterogeneous graph.

\begin{itemize}
    \item \textbf{Nodes, set $\mc{V}$}: users, roles, groups, datastores. Each node has its own handcrafted features. For example, a datastore includes distribution of data types, each user node has an associated risk score.
    \item \textbf{Edges}: a pair of nodes is connected with an undirected edge, if they have participated in the execution of a cloud operation (e.g. reading data, modifying resource configuration, etc) and its weight, $e_{v,u,\tau}$, is proportional to the frequency of such an operation. 
    \item \textbf{Edge types, set $\mc{R}$} carry semantic meaning of executed operations: data flow edges (e.g. transferring data), configuration update edges (e.g. updating configuration of cloud resources) and others.
\end{itemize}

\paragraph{Graph neural network.} We adapted Relational GCN~\cite{r_gcn} approach to heterogeneous graphs, however, we replaced GCN~\cite{gcn} modules with GraphSage~\cite{graph_sage} modules (one per each graph relation type) to achieve \textit{inductive} graph representation learning.
Our GNN-based embeddings are quite general, we successfully use them for multiple downstream tasks such as anomaly detection and data visualization.


Conceptually, our GNN is a superposition of multiple \textit{weighted} GraphSage neural networks where each network operates on a specific relation type $\tau \in\mc{R}$ (Eq.~\eqref{eq:graph_sage_nbd_aggregate_rule}--\eqref{eq:gnn_norm_rule}). 
At each search depth (parameter $k$) nodes aggregate information from their local neighbors, $\mc{N_{\tau}}(v)$, using the \textit{weighted mean} aggregator, into the vector $\textbf{h}^k_{\mc{N_{\tau}}(v)}$, where weights are normalized edge frequencies $e_{v,u,\tau}$ (Eq.~\eqref{eq:graph_sage_nbd_aggregate_rule}).
After that, each node embedding $\textbf{h}^k_{v,\tau}$ gets updated according to the rule~\eqref{eq:graph_sage_update_rule}, where $\textbf{W}^k_\tau$ are trainable weight matrices and
$\oplus$ stands for concatenation of two vectors.
The final node embedding $\textbf{h}^k_v$ at the search depth $k$ (Eq.~\eqref{eq:gnn_update_rule}) is a mean value of normalized relation-specific embedding vectors $\textbf{h}^k_{v, \tau}$ (Eq.~\eqref{eq:gnn_norm_rule}).

The search depth is a design parameter and if it is too high, then a GNN may suffer from over-smoothing. Thus, we set it to 2 because it is sufficient for our experiments.
We use 50-dimensional $\textbf{h}^2_v$ (Eq.~\eqref{eq:gnn_update_rule}) vectors as node embeddings in our experiments.
Note that $\textbf{h}^0_v$ vectors in the Eq.~\eqref{eq:graph_sage_nbd_aggregate_rule},~\eqref{eq:graph_sage_update_rule} are initialized by corresponding node feature vectors.
We train our GNN on link prediction task with a cosine distance between node embeddings for 500 epochs.


\begin{align}
&\textbf{h}^k_{\mc{N_{\tau}}(v)} \xleftarrow{} mean_{e_{v,u,\tau}} (\textbf{h}^{k-1}_{u,\tau} \quad  \forall u \in\mc{N_{\tau}}(v)) \label{eq:graph_sage_nbd_aggregate_rule}\\
&\textbf{h}^k_{v,\tau} \xleftarrow{} \sigma (\textbf{W}^k_\tau \cdot (\textbf{h}^{k-1}_{v,\tau} \oplus \textbf{h}^k_{\mc{N_{\tau}}(v)})) \quad \forall v \in\mc{V} \; \forall \tau \in\mc{R}  \label{eq:graph_sage_update_rule}\\
&\textbf{h}^k_{v, \tau} \xleftarrow{} \textbf{h}^k_{v,\tau} / ||\textbf{h}^k_{v, \tau}||_2 \quad \forall v \in \mc{V}\label{eq:gnn_norm_rule} \\
&\textbf{h}^k_v \xleftarrow{} mean (\textbf{h}^k_{v,\tau} \quad \forall \tau \in \mc{R})  \label{eq:gnn_update_rule}
\end{align}

\begin{figure*}[tbp]
\centering
\begin{minipage}[tbp]{0.33\linewidth}
\centering
\includegraphics[width=\textwidth]{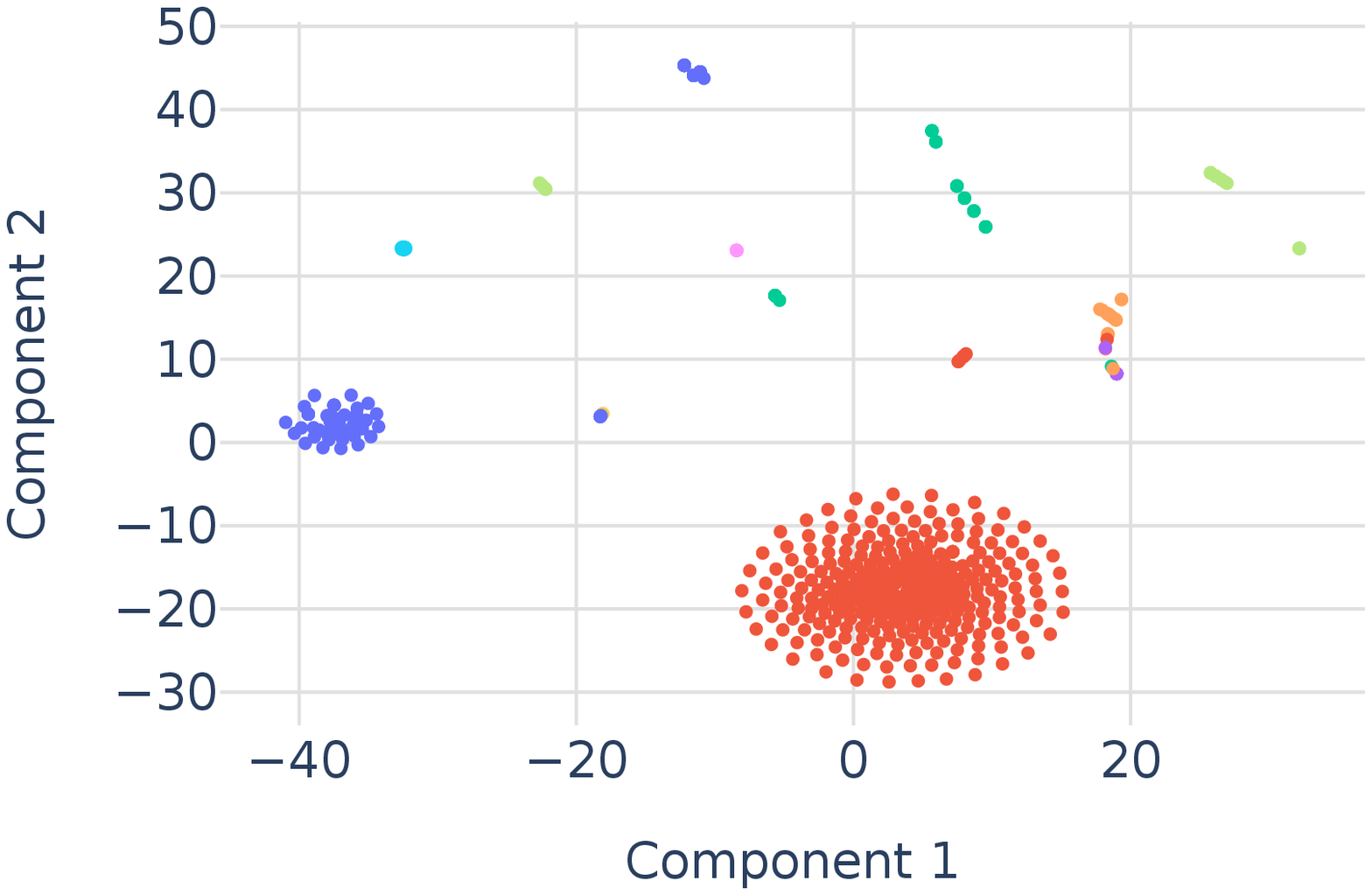}\par
\subcaption{}
\label{fig:gnn_embeddings}
\end{minipage}
\hfill
\begin{minipage}[tbp]{0.33\linewidth}
\centering
\includegraphics[width=\textwidth]{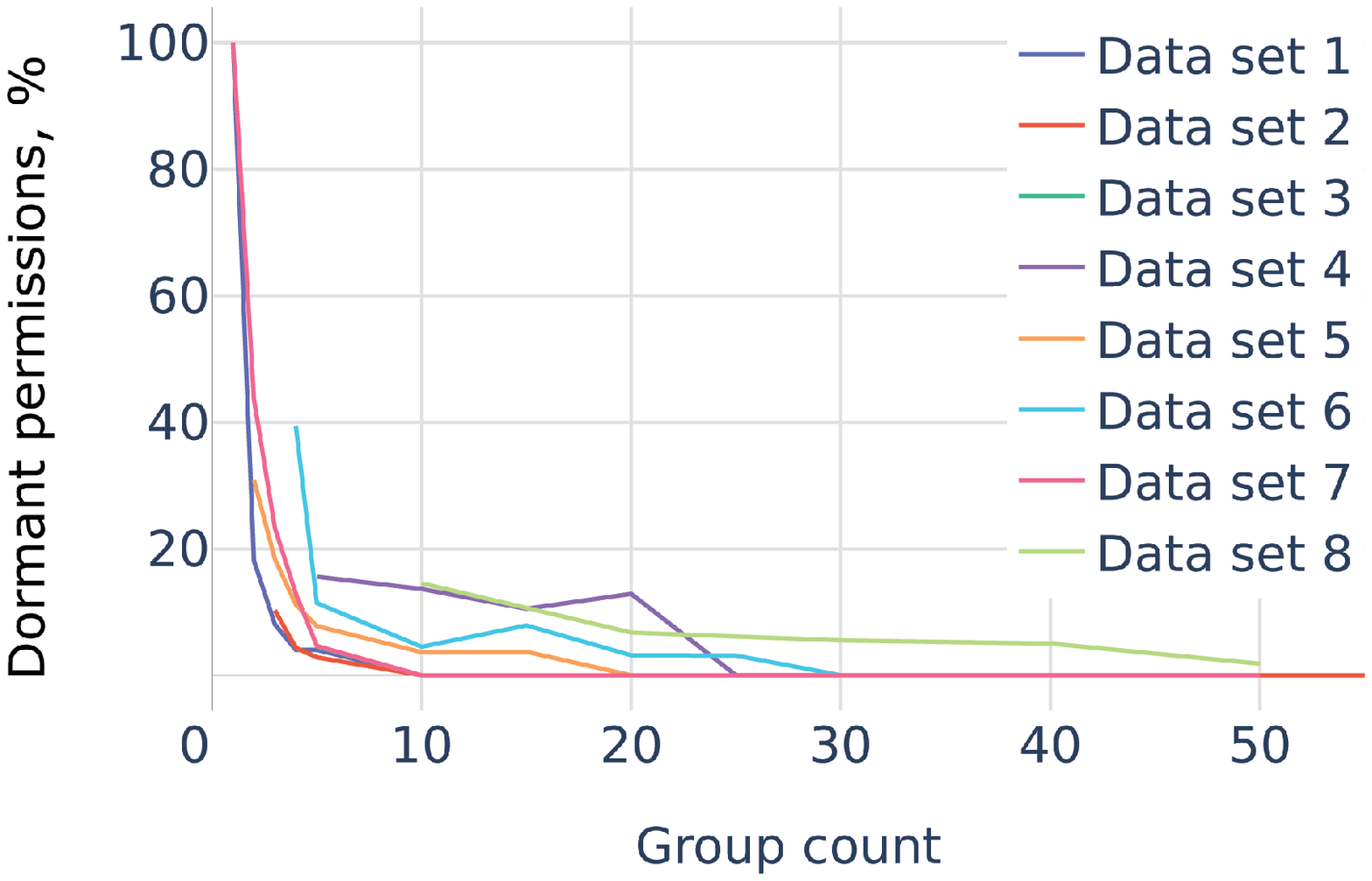}\par
\subcaption{}
\label{fig:rel_dark_perm}
\end{minipage}
\hfill
\begin{minipage}[tbp]{0.33\linewidth}
\centering
\includegraphics[width=\textwidth]{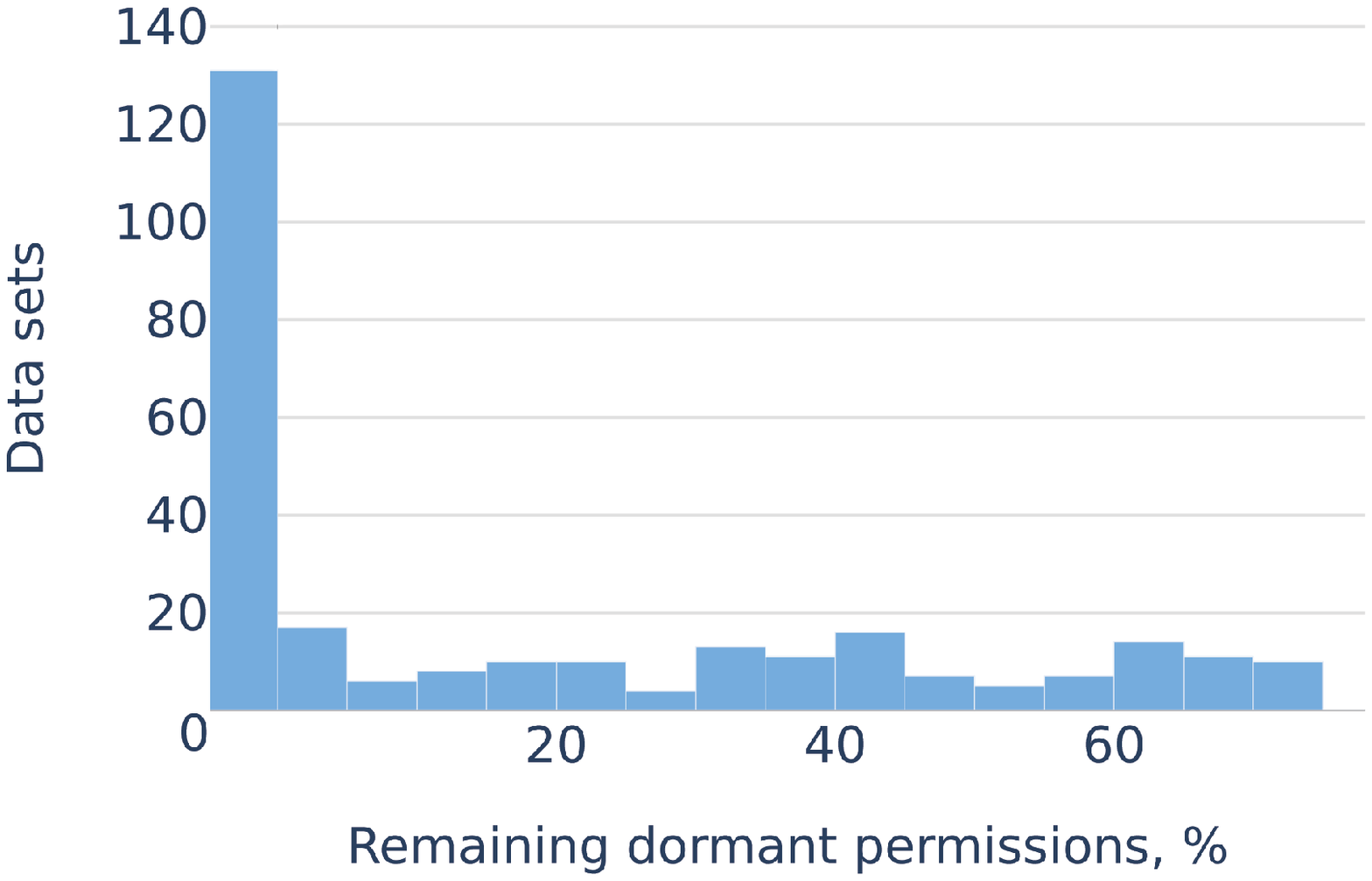}\par
\subcaption{
}
\label{fig:synth_graph_exp}
\end{minipage}

\caption{
        (\textit{a}) T-SNE projection of 50-dimensional Graph-NN identity embeddings (data set 7); 15 distinct clusters corresponding to different job roles are highlighted.
        (\textit{b}) Relative percentage of remaining dormant permissions after IAM policy optimization as a function of the number of generated data access groups. \sysname{} significantly reduces the amount of dormant permissions and often reaches 0\% level with a small number of datastore access groups. For group count more than 50, there are no dormant permissions, therefore not shown in the plot.
        (\textit{c}) To verify the generalizability of our approach, we evaluate \sysname{} on 280 synthetic data sets. In most cases, \sysname{} achieves significant reduction of dormant permissions.
        }
\label{fig:panel_1}
\end{figure*}

\paragraph{User embeddings to CP parameters.} After training is complete we extract user embeddings and cluster them with k-means algorithm. A grid search with k in range of [5, 25] gives us an optimal value of k, $k_{opt}$, that corresponds to a point of maximum curvature (`knee' point) of the function that maps k to k-means' objective value (sum of squared distances of samples to their closest cluster center)~\cite{kneedle}. The range of k encodes our domain-specific knowledge - we expect to identify between 5 and 25 sufficiently different job roles at an organization. Figure~\ref{fig:gnn_embeddings} visualizes 15 clusters detected in one of the data sets. We shared clustering results with organizations and received back a confirmation of good approximation quality of employees' job roles.


GNN embeddings let us define user dissimilarity function $\alpha_{user}(u_1, u_2)$ consistently with GNN design: it is the cosine distance between embeddings of users $u_1$ and $u_2$. The threshold $\alpha$ is the maximal cluster diameter computed as the cosine distance between the most dissimilar users within any cluster when setting the parameter k to $k_{opt}$. Significant entropy reduction (Section~\ref{sec:user_embedding_constraints}, Figure~\ref{fig:entropy_reduction}) of data access groups after iteratively adding homogeneity constraints empirically validates the choice of the parameter $\alpha$. If $\alpha$ was too high, then we would not observe entropy reduction.

\section{Algorithm}
\label{sec:algorithm}

\begin{algorithm}[tb]
\begin{algorithmic}[1]
\State  $\textbf{Input:} \ org's \ cloud \ configuration$, $|\mc{G}|$ \Comment{Section \ref{sec:model}}
\State $\textbf{Output:} \ Hardened \ IAM \ policies$
\\
\State $Train \; Graph \ Neural \ Network \ (GNN)$   \Comment{Section \ref{sec:gnn}}
\State $\alpha_{user}(u_1, u_2), user \ cluster \ assignment \gets GNN$
\State $CP \gets CP_{basic}$  \Comment{Table \ref{tab:cp}}
\State $solution \gets solve(CP)$
\\ 
\If {$solution = \emptyset$}
    \State \Return \textit{Infeasible CP}
\EndIf
\\
\Repeat  \Comment{Section \ref{sec:model}, "Constraint generation"}
    \ForAll{$\tlg \in \mc{G}$}
        \State \textit{\# Identify dissimilar user pair} $(u_1, u_2)$
        \State $(u_1, u_2) \ s.t. \ u_1,u_2 \in \tlg and \ \alpha_{user}(u_1, u_2) > \alpha$
        \State \textit{\# Add additional constraints}
        \State $CP \cup \{\small \ug(u_1, g\prime) + \ug(u_2, g\prime) \leq 1 \ \forall g\prime \in \mc{G}\}$
    \EndFor
    \State $solution \gets solve(CP)$

\Until{$solution \neq \emptyset$}
\\
\State \textit{Hardened IAM poclicies} $\gets$ \textit{extract(last feasible solution)}
\end{algorithmic}
\caption{\sysname{}: high-level description}
\label{alg:alg}
\end{algorithm}

In this section we turn concepts outlined in Sections~\ref{sec:model},~\ref{sec:gnn} into an imperative Algorithm~\ref{alg:alg} to clarify \sysname{}'s workflow. 
After \sysname{} gets deployed in an organization's cloud environment, it starts mining cloud configuration to convert it into a graph representation. Hence, those details lie outside the scope of our paper.
Having computed the graph representation, \sysname{} trains a GNN (Section~\ref{sec:gnn}), which is used to generate graph node embeddings and to cluster users (Algorithm~\ref{alg:alg}, lines 4,5).
At this point, \sysname{} can proceed with solving the basic CP problem outlined in the Table~\ref{tab:cp} (Algorithm~\ref{alg:alg}, lines 6,7).
The basic CP problem may turn out to be infeasible (Algorithm~\ref{alg:alg}, lines 9--11) if the specified number of datastore access groups is insufficient (input parameter $|\mc{G}|$).

If \sysname{} can find a feasible solution to the basic CP problem (Algorithm~\ref{alg:alg}, line 7), then it proceeds with iterative constraint generation (Algorithm~\ref{alg:alg}, lines 13--21).
This process continues until all user dissimilarity constraints get satisfied or the CP problem turns infeasible at some iteration. 
In practice, constraint generation usually stops due to encountering an infeasible CP problem. Also, at each iteration we add mini-batches of constraints rather than individual constraints 
to amortize time needed for running the CP solver.
When the execution reaches the end (Algorithm~\ref{alg:alg}, line 23), it is guaranteed that the feasible solution has been found at either the final iteration or the one before that.
Finally, the CP solution gets decoded into new IAM policies that are concatenated with existing ones using \textit{logical} AND operation to reduce the amount of \textit{dormant} permissions.
\arxiv{\vspace{-0.1in}}

\section{Attack Models}
\label{sec:attack}
Besides reducing the dormant permissions, we now describe different attack models that further test our optimized IAM policies. In \textit{random} attack model, we assume that an attacker does not have information about the internal IAM configuration of an organization. Therefore, the attacker randomly tries to compromise $k$ users. Once compromising $k$ users, the attacker gets access to all the datastores that can be accessible by any of the $k$ users.

In the \textit{worst-case} attack model, we assume that an attacker has full observability of an organization's IAM policies. Therefore, the attacker carefully plans to attack $k$ users such that it can maximize the blast radius (or the number of datastores it can access). 
We show that this problem is NP-Hard; also, it is \textit{monotone} and \textit{submodular}, thus a simple greedy approach provides a constant factor approximation.

\subsection{Formalization of the Worst-case Attack Problem}
The proof of NP-hardness relies on reduction from the set cover problem.
Suppose we have an instance of the Set Cover problem, where the sets are $X_1$ to $X_m$. Each set $X_i$ is composed of elements $u_{i1}$ to $u_{in_i}$ where $n_i$ denotes the number of elements in set $X_i$. Let the total number of elements in the universe be $n$.
We create an instance of IAM problem as next. 

\begin{itemize}

\item We create $m$ users corresponding to each set $X_1$ to $X_m$.

\item We create $n$ datastores, one for each element of the universe.

\item If set $X_i$ contains element $j$, we allow user $X_i$ access to the datastore $j$.
\end{itemize}

We now solve the worst-case attack problem on this datastore access graph by choosing $k$ users to compromise. If the number of datastores accessible is the same as the universe, then the set cover problem has a solution with $k$-cover, otherwise not. 

To properly simulate the worst-case attack, which is NP-hard, we have to uncover its submodular nature and use an appropriate approximation.

\paragraph{Monotonicity.} Let $S$ be the set of users being compromised. Let $f(S)$ denotes the number of datastores that become accessible as a result. The function $f$ is monotone. Let $u\notin S$ be another user. $f(S\cup \{u\}) \geq f(S)$ as the number of accessible datastores cannot decrease if an additional user is compromised.

\paragraph{Submodularity.} The function $f$ is also submodular. Let $S'\subseteq S$, and let $u$ be a user not in $S$. Then, we must show:
\begin{align}
    f(S\cup \{u\}) - f(S) \leq f(S'\cup \{u\}) - f(S')
\end{align}

Let us consider the expression $ f(S\cup \{u\}) - f(S)$. It will count those datastores that can only be accessible by user $u$ and none of the users in the set $S$. Let us denote this number as $n_{u\vert S}$; $n_{u\vert S^\prime}$ is defined analogously. We must have $n_{u\vert S} \leq n_{u\vert S^\prime}$ because $S^\prime \subseteq S$. This is becuase the additional access to datastores granted by the user $u$ over datastores accessible by $S$ must be smaller than the additional access granted by the user $u$ over datastores accessible by $S^\prime$.

\paragraph{Constant factor approximation.} The standard greedy algorithm that iteratively selects the user $u$ that provides the maximum marginal gain in terms of the function $f$ is guaranteed to provide $(1-1/e)$-approximation~\cite{Nemhauser1978}. Using this approach, we can approximately select $k$ users to compromise for simulating the worst-case attack.

\arxiv{\vskip{-2pt}}
\section{Experimental Setup}

\begin{figure*}[tbp]
\centering
\begin{minipage}[tbp]{0.24\linewidth}
\centering
\includegraphics[width=\textwidth]{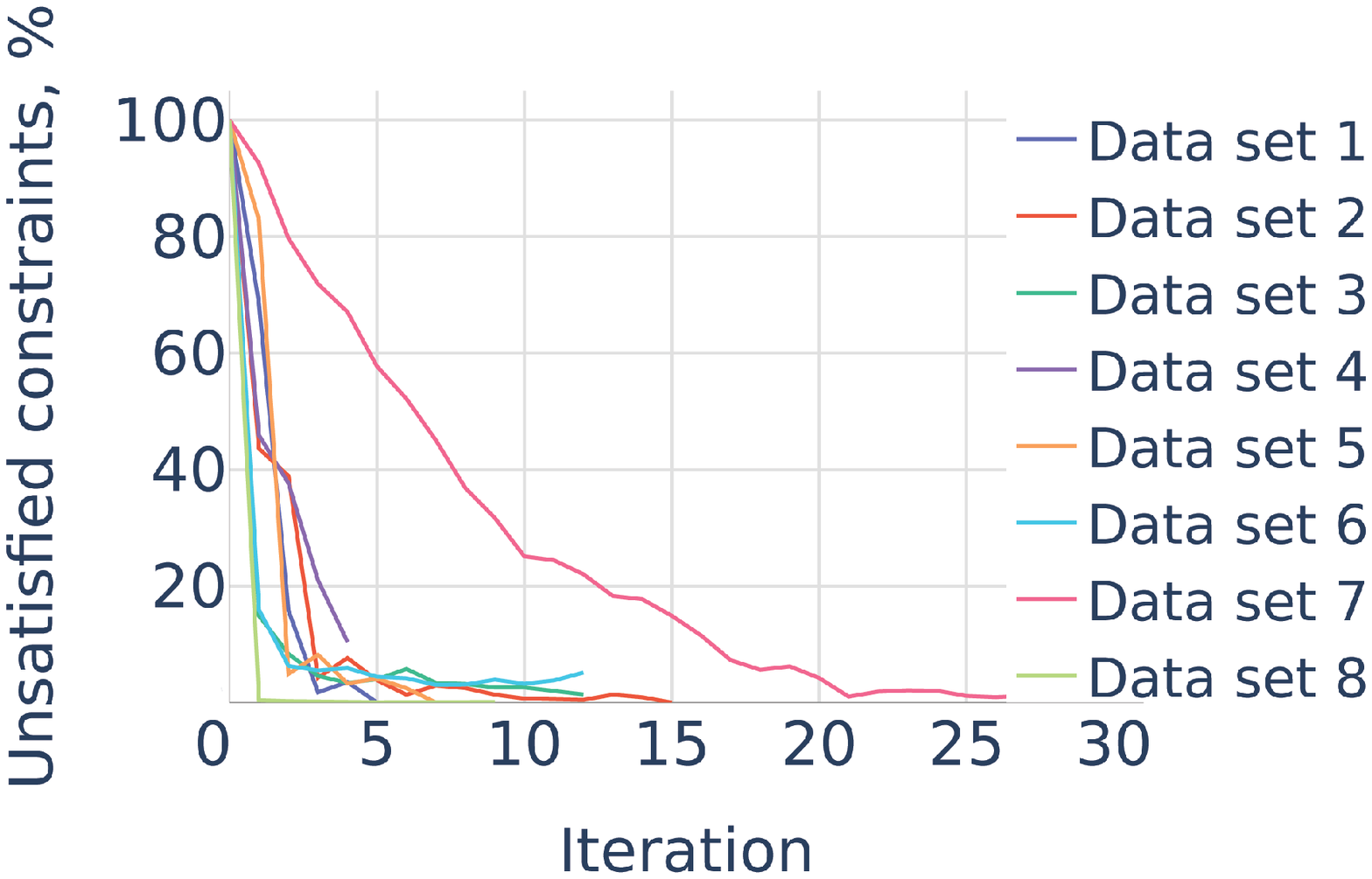}\par
\subcaption{}
\label{fig:unsatisfied_constraints}
\end{minipage}
\begin{minipage}[tbp]{0.24\linewidth}
\centering
  \includegraphics[width=\textwidth]{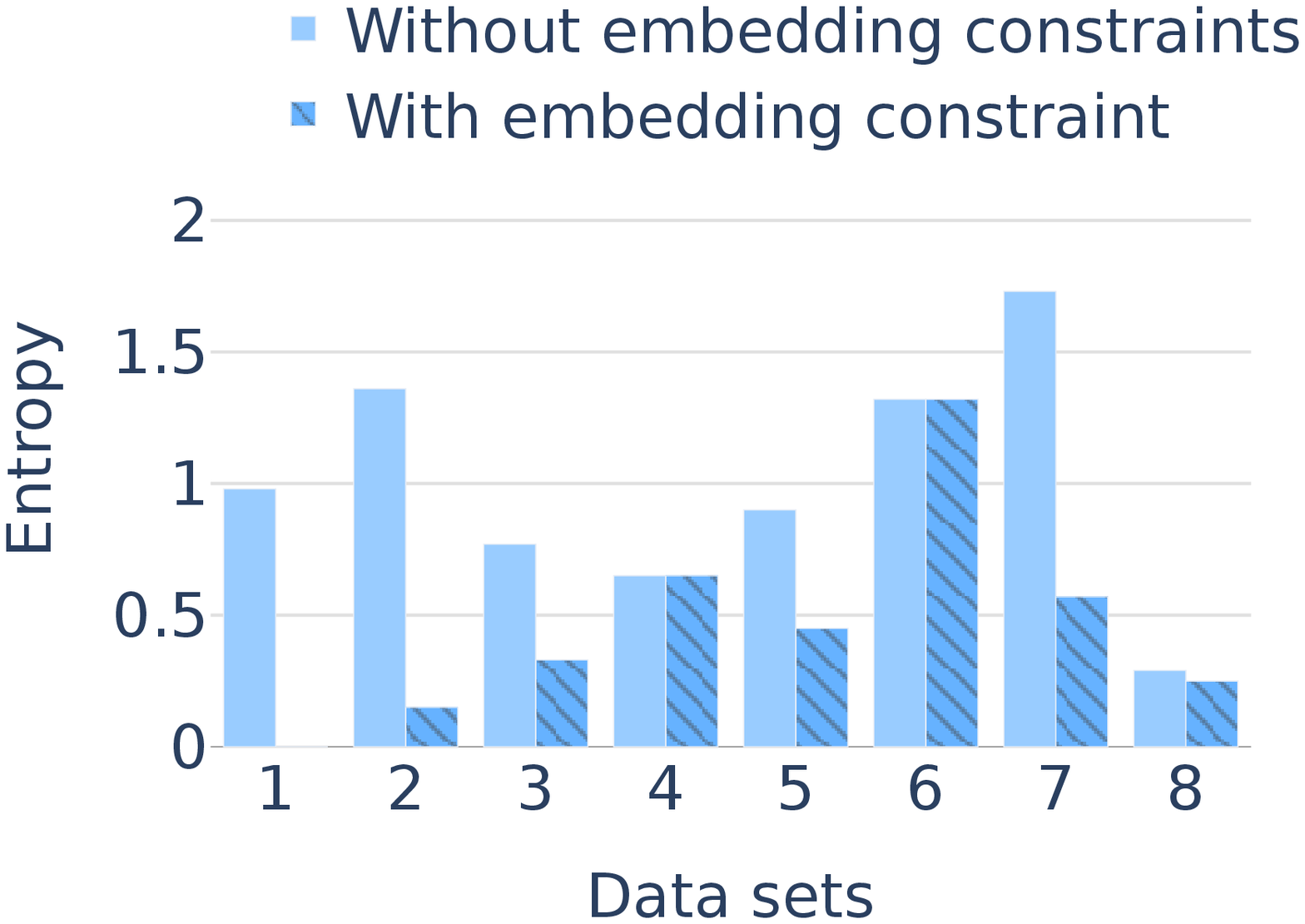}\par
  \subcaption{}
\label{fig:entropy_reduction}
\end{minipage}
\begin{minipage}{0.24\linewidth}
\centering
\includegraphics[width=\textwidth]{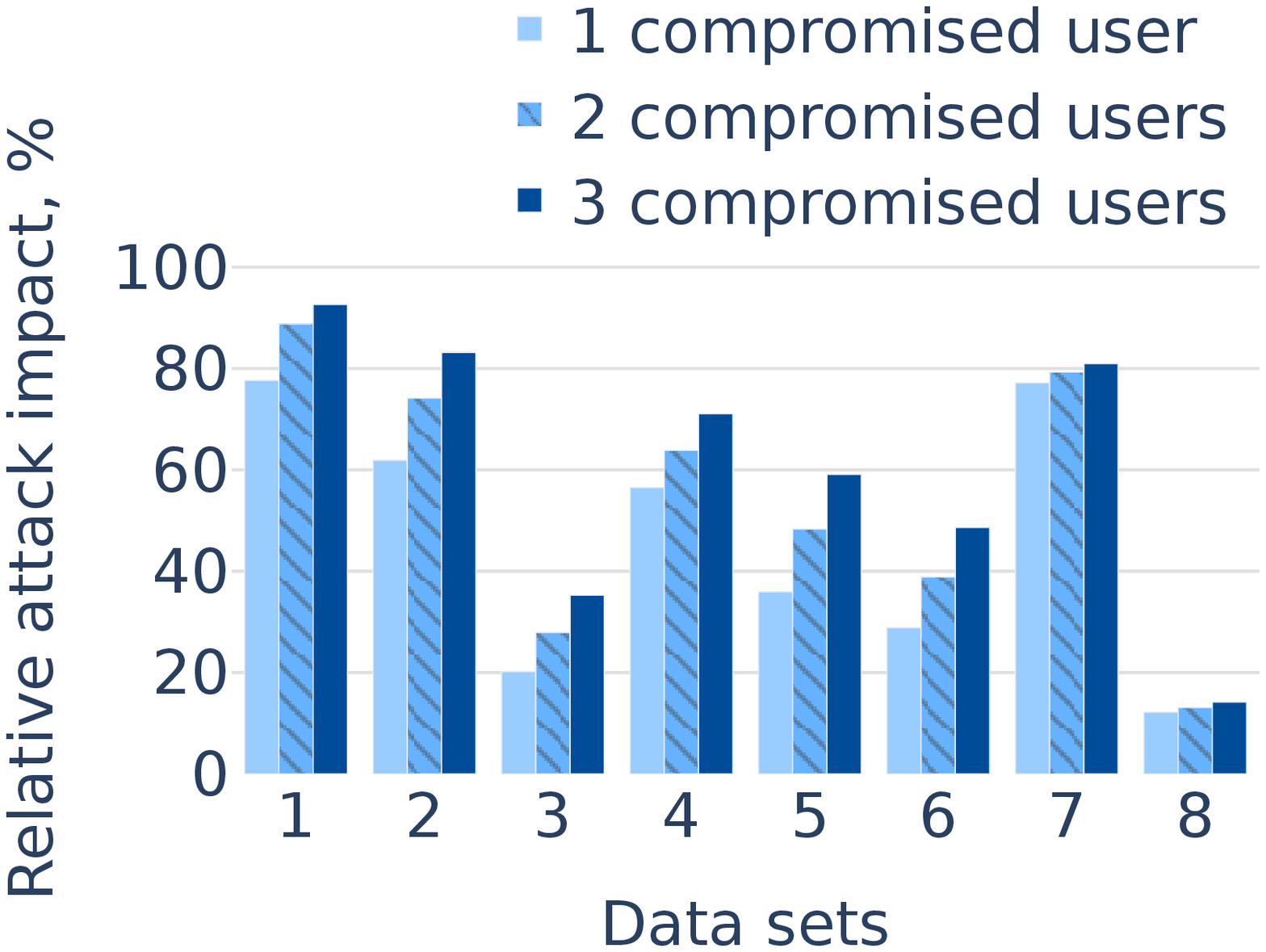}\par
\subcaption{}
\label{fig:avg_attacker_impact}
\end{minipage}
\hfill
\begin{minipage}{0.24\linewidth}
\centering
  \includegraphics[width=\textwidth]{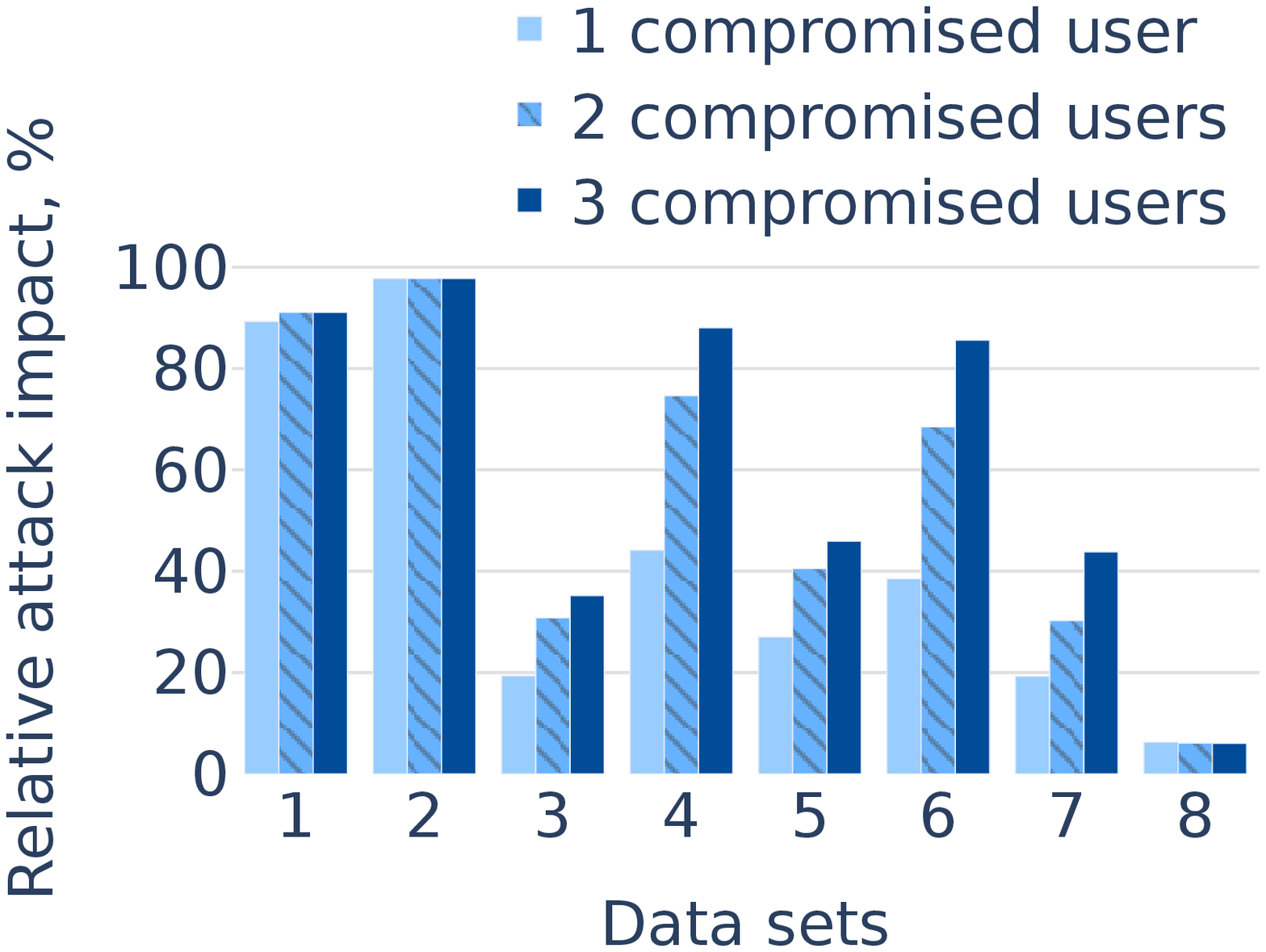}\par
  \subcaption{}
\label{fig:max_attacker_impact}
\end{minipage}

\arxiv{\vspace{-0.1in}}
\caption{
        (\textit{a}) Cumulative fraction of satisfied user embedding constraints at each iteration.
        (\textit{b}) Entropy of generated solutions before and after adding user embedding constraints.
        (\textit{c, d}) Attack impact (lower better) on an organization under an assumption of 1-3 compromised users.
        (\textit{c}) The average-case impact considering an attacker possesses incomplete information about IAM policies.
        (\textit{d}) The worst-case impact considering an attacker possesses perfect information about IAM policies.
        }
\label{fig:panel_2}
\end{figure*}

\begin{table}[htbp]
\small 
\begin{center}
\begin{tabular*}{0.46\textwidth}{ |p{0.1\linewidth} | p{0.1\linewidth} | p{0.15\linewidth} | p{0.18\linewidth} | p{0.178\linewidth}| }
 \hline
Graph & Users & Data stores & Dynamic edges & Permission edges\\
 \hline
 1 & 13 & 56 & 513 & 1969 \\ 
 2 & 32 & 89 & 1192 & 16791 \\ 
 3 & 39 & 341 & 1602 & 28880 \\ 
 4 & 57 & 572 & 7153 & 37854 \\ 
 5 & 60 & 88 & 757 & 4115 \\ 
 6 & 64 & 258 & 2418 & 72272 \\ 
 7 & 112 & 163 & 2789 & 4025 \\ 
 8 & 150 & 600 & 3095 & 11517 \\ 
 \hline
\end{tabular*}
\end{center}
\arxiv{\vspace{-0.1in}}
\caption{Properties of real world graph instances}
\label{table:data_sets}
\end{table}

We evaluate \sysname{} on both synthetic and real world instances. Each real world instance represents cloud infrastructure within one or more departments at a commercial organization. 
In addition to 8 real world data sets, we generated 280 synthetic data sets using real data sets as baselines.
Throughout the paper we mostly focus on real world data sets and use synthetic data only for evaluation of statistical effectiveness of the proposed approach.
Independent security experts verified correctness and interpretability of generated IAM policies.
Moreover, generated policies are correct by design (Section~\ref{sec:model}).
The section is organized as follows: first we show effectiveness of our basic CP formulation, then we demonstrate the effect of adding constraints based on GNN embeddings, and  finally we conclude with attack simulations.

\paragraph{CP Versus MILP.} Even though our CP formulation can be solved with an MILP solver after being linearized, we used IBM CP solver~\cite{ibm_cp} because linearization of non-linear functions (e.g. max, logical operators, etc) introduces a large number of additional variables, thus making the problem unsolvable in a practical amount of time. 
In the case of the largest instance (instance \#8) linearization increases the number of variables by $\sim$32 times and the number of constraints by $\sim$9 times. Specifically, MILP formulation contains 3,900,750 variables and 4,280,130 constraints vs 120,000 variables and 472,801 constraints in the CP case (not counting homogeneity constraints). 
As a consequence, IBM MILP solver failed to find any feasible solution within 2.5 hours.
Across all 8 real data sets, the size of CP problem lies within the range of 2,108--196,000 variables and 6,528--300,000 constraints (after incorporating homogeneity constraints).
\sysname{} is designed to be used as a decision support tool by system admins, thus we prioritize fast solving time by setting the time limit to 15 minutes.
All experiments were conducted on AWS c5.24xlarge virtual machine equipped with 96 vCPU and 192 GB of RAM.

\paragraph{Real world graphs.} Real world graphs were shared by IT departments of 8 commercial organizations (Table~\ref{table:data_sets}).
Graphs are very sparse as we would expect in a security setting - only certain accesses (permission edges) are allowed. However, most of them remain unused: the number of \textit{dynamic edges} that represent actual data accesses is even smaller. Densities of graphs built on permission and dynamic edges differ by 1.4 - 29.9 times. High coefficients correspond to poorly designed (over-priviledged) security models. The number of users varies between 13 and 150, while the number of datastore nodes lies within the range of 56 - 600.

\paragraph{Synthetic graphs.} We use real graphs as baselines to generate 280 synthetic graphs. 
We vary the number of nodes, but keep graph density as is, i.e. in the range of 0.259 $\pm$ 0.198 (avg $\pm$ std).
To generate a synthetic graph, we first sample the number of users and datastores from uniform distributions over the following intervals [10, 150] and [50, 300] respectively that cover variations of those parameters across real graphs. We deliberately set the maximum number of data stores fewer than 600 (instance 8) to speed up computations. After fixing node counts we sample with replacement the actual nodes from a real world graph, which is chosen at random. Then we add Gaussian \textit{N(0, 0.01)} noise to node embeddings and renormalize them. To match the graph density with the density of the underlying baseline we sample edges from a multinomial distribution, where each component is proportional to the cosine distance between a user and a datastore embeddings. Also we enforce the invariant that dynamic edges are always a subset of all permission edges. A synthetic graph generated in such a way is an "upsampled" version of an underlying real world graph.


\subsection{Reduction of Dormant Permissions}
\paragraph{Real world instances.} \sysname{} significantly reduces \textit{dormant} permissions (Figure~\ref{fig:rel_dark_perm}) over the current IAM policy of companies. For this purpose we vary the number of datastore access groups $|\mc{G}|$ from 1 up to {100} with an increment of 5. 
The solver's time limit is fixed at 15 minutes. 
When the number of datastore access group is too small, the problem often becomes infeasible.
For most data sets, the fraction of remaining \textit{dormant} permissions quickly goes down to 0\% as we increase the number of datastore access groups. The only exception is the instance 8 (the largest data set), which requires 50 data access groups. According to these results we set the number of data access groups to 20 for instances 1-7, and 40 for the instance 8 in other experiments. We also highlight that it is important to have as few as possible datastore access groups to keep IAM policy interpretable. Our results show that even with additional 5 groups, dormant permissions reduce significantly over the existing IAM policies.



\paragraph{Synthetic experiments.} To evaluate \sysname{} across larger number of instances, we use 280 synthetic graphs. Figure~\ref{fig:synth_graph_exp} shows unnormalized distribution of the remaining dormant permissions in a graph while running the solver for at most 15 minutes, and fixing the datastore access groups at 20. Each bin spans 5\% interval, graphs with no \textit{dormant} permissions (0\%) fall into the very first bin. The number above a bar denotes how many instances fall in the corresponding interval. 
The histogram is skewed towards the left-hand side: in 46.7\% of cases \sysname{} reduces dormant permissions down to 0\%. However, we observe some outliers where the solver is unable to reach dormant permissions 0\% level. This mostly happens because of either exceeding the time limit or the need for a larger number of data access groups due to instance complexity. These results can be improved by setting a higher time limit for the CP solver or increasing the number of data access groups.  

\subsection{User Embedding Constraints}
\label{sec:user_embedding_constraints}
To make generated IAM policies interpretable we follow constraint generation approach outlined in Section~\ref{sec:gh}, which leads to generating mostly homogeneous data access groups.

Figure~\ref{fig:unsatisfied_constraints} shows the fraction of unsatisfied embedding constraints at each iteration when setting the number of data access groups to 20. Depending on the data set, it takes 10-20 iterations to satisfy more than 95\% of embedding constraints.
To verify that such an approach produces mostly homogeneous groups, we compare the entropy of generated data access groups with respect to users' cluster assignment (Figure~\ref{fig:entropy_reduction}) before and after adding embedding constraints. In all cases except the data sets 4 and 6, embedding constraints drastically reduce group entropy. In the case of the data sets 4 and 6, the problem becomes infeasible at the very first constraint relaxation iteration.




\subsection{Simulated Security Attacks}
We consider two security attacks that fundamentally differ in terms of knowledge that an attacker posses. In both cases an attacker tries to compromise $k$ users, where $k \in \{1,2,3\}$. An attack's impact is the number of datastores that an attacker can get access to. Hence, we report the relative attack impact, which is the the ratio between the number of compromised datastores after applying \sysname{} and the number of compromised datastores in the existing cloud infrastructure. The lower ratio is, the more noticeable effect \sysname{} has on the IAM policy optimization. We notice that \sysname{}'s effect can be masked by high-degree user nodes, especially, in the worst-case attack because the greedy algorithm keeps selecting such nodes. High-degree user nodes impose a severe security risk on an organization and this issue should be mitigated using traditional software engineering methods - splitting nodes into multiple nodes of lower degrees.
To evaluate \sysname{} in the worst-case attack scenario, we removed top-30\% of high-degree nodes.

If an attacker has no information about the implemented IAM policies, then we can estimate the average impact of compromising $k$ user identities (Figure~\ref{fig:avg_attacker_impact}). For example, in the case of the instance 8 \sysname{} achieves the highest impact reduction (86\% - 88\%) even though it is the largest real world data set. However, impact reduction is minimal for the instance 1 (8\% - 22\%) due to the presence of high-degree user nodes.

If an attacker has a perfect knowledge of the IAM policies, then the attacker can cast the problem to max k-cover and solve it using a greedy approximation algorithm. Figure~\ref{fig:max_attacker_impact} illustrates such a case study. \sysname{} minimizes the worst case attack impact for organizations 3, 5, and especially 8 by 41\% - 89\%. However, organizations 1 and 2 remain almost unaffected due to the large number of high-degree nodes remaining in the data set after removing the top-30\% of user nodes sorted by their degrees.


Our current simulated attacks are primarily based  on the interaction patterns between users and datastores, and the observations available to an attacker. We note that using the rich language of constraint programming, we can also generate \textit{automated} attacks that achieve malicious goals while complying with IAM policies. We leave this as part of the future work.

\arxiv{\vskip{-2pt}}
\section{Conclusion and Future Work} 
Given the increasing popularity of cloud computing, associated security issues are also increasing in severity. We developed a principled approach for the key problem of optimizing IAM policies to reduce the attack surface of an organization’ cloud setup (or the so-called \textit{dormant} permissions that allow users access datastores which are not needed for users' business functions). We presented a formulation of IAM policy optimization using constraint programming and graph representation learning, identified key constraints which IAM policies should satisfy, and then tested the resulting IAM policies on 8 real world and multiple synthetic data sets. Our results show that \sysname{} is highly effective in reducing \textit{dormant} permissions and generating interpretable IAM policies. Our framework also opens the door to the application of a host of AI methods to address additional security problems in cloud infrastructure.

Several promising opportunities exist for enriching IAM policy generation with more complex constraint types. For example, temporal constraints that define the order of datastore accesses and set constraints that define what subset of automated services can simultaneously access a given resource can significantly limit the range of attackers' actions after they compromise a cloud environment.
To formulate such constraints, we can use program invariants inferred during the program analysis of automated services.
Moreover, violations of such constraints at runtime can also be used for anomaly detection.


\small 
\bibliographystyle{named}
\bibliography{ref}
\end{document}